# Holeum, Enigmas of Cosmology and Gravitational Waves

L. K. Chavda
And
Abhijit L. Chavda
Chavda Research Institute
49 Gandhi Society
City Light Road
Via Parle Point
Surat 395007
Gujarat, India.
E-mails: lk_chavda@rediffmail.com and al_chavda@rediffmail.com

June 17, 2004.

ABSTRACT

The principle of nuclear democracy is invoked to prove the formation of stable quantized gravitational bound states of primordial black holes called Holeums. The latter come in four varieties: ordinary Holeums H, Black Holeums BH, Hyper Holeums HH and the massless Lux Holeums LH. These Holeums are invisible because the gravitational radiation emitted by their quantum transitions is undetectable now. The copiously produced Holeums form an important component of the dark matter (DM) and the Lux Holeums an important component of the dark energy (DE) in the universe. A segregation property puts the Holeums mainly in the galactic haloes (GH) and the domain walls (DW) explaining the latters' invisibility now. Cosmic rays (CR) are produced by two exploding black holes created in a pressure-ionization of a stable Holeum. Our prediction that more CRs will be emitted by the haloes than by the discs of galaxies already has a strong empirical support. The concentration of the Hs and HHs in the GHs would lead to the formation of Holeum- stars emitting CRs and the gravitational waves (GW). Innumerable explosions of BHs at the time of decoupling of gravity from the other interactions lead to inflation and baryon asymmetry (BA). A substantial cosmic background of matter and GWs and an infra-quantum gravity (infra-QG) band and an ultra-QG band of GWs and their emission frequencies are predicted. A unique quantum system containing matter-energy oscillations similar to the pyrgons in higher dimensional Kaluza-Klein theories is found.

I. INTRODUCTION

In the last century there have been three great droughts in physics. These are the atomic drought, the electroweak drought and the Quantum Gravity (QG) drought. The first two have been overcome and now we are facing the third and the longest one. Since the discovery of quantum theory attempts have been going on to quantize the gravity but to no avail. Therefore it is instructive to examine how the first two were tackled. Whenever one is faced with a new phenomenon one tries to understand it, in the first instance, in terms of the existing theory, however incomplete and rudimentary it may be. This may be called the method of the Standard Paradigm (SP). Historically this has been very successful. Bohr and Glashow successfully predicted the spectrum of the hydrogen atom and the Charm quark, respectively, by the method of SP. They achieved important breakthroughs which eventually led to the correct theories. In reference [1] and in this paper we follow the same method to tackle the problem of the bound states of the Primordial Black Holes (PBH). We call it the Standard Paradigm of Quantum Gravity (SPQG). We use the Newtonian Gravity (NG) and the Schrodinger equation supplemented by the so-called Bound State Wisdom (BSW) culled from three layers of matter. It holds that only the asymptotic behaviour of the binding potential is sufficient to guarantee the order of magnitude correct values of the bound state parameters. Just as the discovery of the asymptotic freedom enabled the particle physicists to use the nonrelativistic Schrodinger equation to describe the bound states of the strongly interacting quarks, our discovery of the BSW enables us to investigate the properties of the bound states of the PBHs called the Holeums using the NG despite the fact that the Holeums were produced in the strong field regime of the General Relativity (GR). In reference [1] this led to plausible resolution of three enigmas of cosmology, namely, Dark Matter (DM), Domain Walls (DW) and the invisible Galactic Haloes (GH) in terms of the Holeums. The atomic transitions of the Holeums emit Gravitational Waves (GW) which may be detected by the detectors like the Laser Interferometer Gravity wave Observatory (LIGO). In this paper we extend the list of successes of the SPQG to include Dark Energy (DE), the origin of the Cosmic Rays (CR), the Inflation of the universe, the Baryon Asymmetry (BA) and a resolution of the conflict between the Hawking theorem and the Principle of Nuclear Democracy (PND). The latter may be stated as follows: No submicroscopic primordial particle is more or less fundamental than the others. Now we will prove that the PND implies the existence of stable Holeums. We put it in the form of the following theorem.

Dark Matter Existence Theorem: The PND implies the existence of the stable gravitational bound states of PBHs called the Holeums which may constitute an important component of the Dark Matter in the universe.
Proof: A vast quantity of PBHs was produced as a result of the density fluctuations in the early universe. However, according to the Hawking's theorem all the PBHs got explosively evaporated away right at their births. If so, they have had no role to play in the constitution and the evolution of the universe. In contradistinction to this, all the other primordial particles such as the quarks, the leptons and the gauge bosons, etc. have had important roles to play in the universe. This means that the PBHs are less fundamental than the others. This is a violation of the PND. It also means that nature produced a vast quantity of PBHs merely to fritter it away. Since nature is neither frivolous nor spendthrift this creates an enigma which we call the role enigma for the PBHs. The only way to resolve this enigma and to avoid the conflict with the PND is to postulate the existence of stable bound states of PBHs which we have called the Holeums. The atomic transitions of the latter emit gravitational radiation that we cannot detect as yet. Hence the Holeums form a part of the DM. Since a vast quantity of PBHs was produced in the early universe the Holeums may constitute an important component of the DM in the universe today. Hence the theorem.

We can arrive at nearly the same result as follows. Consider the following facts. (1) The presence of a vast quantity of the PBHs in the early universe. (2) The strength of the gravitational interaction is greater than or equal to that of the unified strong-electroweak interaction above the unification temperature, $T_u$, say. (3) The rate of gravitational interaction is greater than that of the expansion of the universe above $T_u$. (4) The gravitational interaction is always attractive. These facts necessarily imply the formation of the Holeums among the PBHs.

Two examples from nature are very relevant here. (1) A free neutron decays but a neutron in a stable nucleus never does. By analogy a black hole in a stable Holeum will not decay. And we have already proved that a Holeum is stable. There is another reason. If the bound state radius of a Holeum is greater than its Schwarzschild radius then the Holeum is not a black hole. Therefore it will not radiate for the same reason that a hydrogen atom will not. (2) Our second example is a quark. A semi-free quark tends to emit a jet of hadrons but a bound quark does not. The first example shows that unstable particles may form absolutely stable bound states and the second one shows that some primordial particles exist only in bound states. The black holes in Holeums have both of these properties.

Now we would like to give teeth to the principle of nuclear democracy (PND) by predicting the cross section and the reaction rate among the PBHs at the unification temperature $T_u$. The cross section $\sigma$ for the gravitational interactions among the PBHs may be written as

$$\sigma = \alpha_g^2 \left( k T_u / m_P c^2 \right)^\lambda \left( \hbar / m_P c \right)^2 \tag{1}$$

where $\lambda$ is a parameter to be determined, k is the Boltzmann constant and $m_P$ is the Planck mass given by

$$m_P = \left( \hbar c / G \right)^{1/2} \tag{2}$$

and $\alpha_g$ is the gravitational fine structure constant given by [1]

$$\alpha_g = \left( m / m_P \right)^2 \tag{3}$$

For the relativistic domain $mc^2 \ll k T_u$. The rate of gravitational interactions among the PBHs is given by

$$\eta = c \sigma \left( k T_u / \hbar c \right)^3 \tag{4}$$

where $\left( k T_u / \hbar c \right)^3$ is the number density of the PBHs in the relativistic domain. Now the rate of the expansion of the universe is given by the Hubble constant H which is given

$$H = \left( 8\pi^3 / 45 \right)^{1/2} (k T_u)^2 / (\hbar m_P c^2) \tag{5}$$

The PND requires that

$$\eta \geq H \tag{6}$$

In the following we will determine the critical value of $\lambda$ for which $\eta = H$. It is given by

$$\lambda = -1 - (½) \ln[(4\pi^3/45) / \alpha_g^2] / \ln( m_P c^2 / k T_u ) \tag{7}$$

Substituting equation (7) into equation (1) we get

$$\sigma = ( m_P c^2 / k T_u ) ( 4\pi^3 / 45 )^{1/2} ( \hbar / m_P c )^2$$
$$= 6.1320 \times 10^{-67} \text{ m}^2 \tag{8}$$

And the rate of the gravitational interactions among the PBHs is given by

$$\eta = 4.3583 \times 10^{37} \text{ s}^{-1} \tag{9}$$

In view of equation (6) the last two results are the lower bounds. These results are perhaps the first quantitative predictions of the PND. And as they are model-independent they provide an invaluable insight into the conditions prevalent in the early era of the bigbang universe.

This paper contains two important points of departure from reference [1]. (1) Strong arguments were presented in reference [1] to make the formation of Holeums plausible. Here we have proved the latter in two different ways on very general grounds independent of any models. (2) We have replaced the condition $mc^2 \ll k_B T$ by a much weaker one: "The rate of gravitational interactions among the PBHs is greater than or equal to that of the expansion of the universe above the unification temperarure". (3) In reference I it was assumed that a Holeum would be stable if its constituent PBHs did not overlap. That is, we assumed $r_n > 2R$ as the condition for stability of a Holeum. Here $r_n$ is the radius of the bound state with the principal quantum number n and R is the Schwarzschild radius of the two identical black holes constituting the Holeum. However, this criterion is naïve. Therefore in this paper we replace it by equations (17) and (18) given below. These equations follow from the following definition of a black hole. If a bound state has a radius less than or equal to its Schwarzschild radius then the bound state is a black hole otherwise it is not. Surprisingly this leads to only two to eight percent changes in the values of the parameters of the Holeums given in reference I. On the other hand it leads to three new types of Holeum and allows us to resolve the enigmas of DE, CRs, Inflation and the BA in terms of the Holeums.

II. CLASSES OF HOLEUMS.

Let $r_n$ and $M_n$ be the radius and the mass of the bound state with the principal quantum number n. Now it was shown in reference I that

$$M_n = 2m( 1 - \alpha_g^2 / (8n^2) ) \tag{10}$$
$$r_n \approx \pi^2 n^2 R / ( 8 \alpha_g^2 ) \tag{11}$$

where in the latter equation we have assumed $n \gg 1$. From the last two equations it follows that

$M_n / r_n \approx ( 1 – 32 X_1 X_2 / \pi^2 )(c^2/2G)$ (12)

$X_1 = x / 4 - 1 + \Delta$ (13)

$X_2 = x/4 - 1 – \Delta$ (14)

$\Delta = ( 1 - \pi^2 /32 )^{1/2} = 0.831610$ (15)

$x = \alpha_g^2 /n^2$ (16).

Now we adopt the criterion that a bound state is a black hole if

$M_n / r_n \geq c^2/2G$ (17)

And that it is not a black hole if

$M_n / r_n < c^2/2G$ (18)

That is, the bound state is a black hole if its radius is less than or equal to its Schwarzschild radius otherwise it is not a black hole. Now from equation (10) it is obvious that bound states are possible only if

$0 < \alpha_g < 8^{1/2}$ (19).

On the basis of the criterion presented in equations (17) and (18) the interval $(0, 8^{1/2})$ of $\alpha_g$ given in equation (19) can be split into three intervals as follows: We define the region b as

$0 < \alpha_g < 2(1 – \Delta)^{1/2}$ (20)

For this region where $X_1 < 0, X_2 < 0$ one can show that

$(4(1 – \Delta))^{-1/2} \alpha_g < n < \infty$ (21)

We define the region c as

$2(1 – \Delta)^{1/2} \leq \alpha_g \leq 2(1 + \Delta)^{1/2}$ (22)

where $X_1 X_2 \leq 0$ and the region a as

$2(1 + \Delta)^{1/2} < \alpha_g < 8^{1/2}$ (23)

where $X_1 > 0, X_2 > 0$ and a special point $a_+$ by

$\alpha_g = 8^{1/2}$ (24).

Now it can be shown that all the states n= 1, 2, ….. ∞ in the region b, equation (20), satisfy the criterion given in equation (18). That is, none of them is a black hole. We will call them the ordinary Holeums H. They are as stable as a hydrogen atom. Now the ground states n=1 in the region c satisfy equation (17). That is, they are black holes. We will call them the Black Holeums BH. Similarly all the ground states n=1 in the region a satisfy equation (18) just like those in the region b. But whereas $X_1 < 0, X_2 < 0$ for the region b here in the region a we have $X_1 > 0, X_2 > 0$. Therefore the ground states in the region a are also non-black-hole states. However, since their constituent masses are all greater than $m_P$ we will call them the Hyper Holeums HH. Finally, the ground state at the special point $a_+$ is massless as can be seen from equation (10). It will start moving at the speed of light as soon as it is produced. Therefore we will call it the Lux Holeum LH. In summary, we have four classes of Holeum H, BH, HH and LH given in the increasing order of their constituent masses.

III. HOLEUMS AND COSMIC RAYS.

The region b has been exhaustively studied in reference [1]. However, there are two important differences from the latter. The upper limit of the region b given in equation (20) is given by

$m_c = (4(1 – \Delta))^{1/4} m_P$ (25)

whereas in reference [1] it was $m_c = (\pi^{1/2} /2) m_P$. If we expand $\Delta$, equation (15), in powers of $\pi^2/32$ and keep only the first two terms in equation (25) then we get the same result as the one in reference [1]. That is, the present correct value is about 2 % larger. Secondly substituting from equation (21) into equation (11) we get

$r_n > [ \pi^2 /(64(1 – \Delta))] (2R)$ (26).

In reference [1] the corresponding result was $r_n > 2R$. Once again if we expand $\Delta$ as indicated above we get back the old result. The numerical value of the quantity in the square brackets in equation (26) is 0.9158 which is about 8.4 % smaller than unity. The other results of the reference [1] remain unchanged.

A Holeum is a gravitational analogue of a hydrogen atom. It is as stable as the latter. However, it is well known that hydrogen undergoes pressure ionization in large concentrations of the gas such as the stars, galaxies, nebulae etc. The same thing happens to the Holeums, too. Let us calculate the ionization energy of a Holeum. From the equation for the energy eigen-value $E_n$ given in reference [1], it can be shown that the ionization energy of a Holeum having two identical constituents of mass $10^{14}$ GeV / $c^2$ is only 250 eV. This corresponds to a temperature of $2.5 \times 10^6$ K which readily obtains in stars and galaxies. It is clear that for constituent masses smaller than $10^{14}$ GeV / $c^2$ the ionization energy will be even smaller. However, the binding energy of a Holeum having constituent masses $10^{15}$ GeV/ $c^2$ or greater is 25 MeV or greater which corresponds to the temperature greater than $10^{11}$ K which is far higher than that available in stars and galaxies. It is clear that the pressure ionization of Holeums having constituent masses upto $10^{14}$ GeV / $c^2$ is quite possible in the galaxies and other large concentrations of Holeums such as the DWs, if the latter exist. The pressure ionization of a Holeum will produce two individual black holes which will explode instantly into two showers of particles. These are the CRs that we receive on the Earth. Each shower will have a total energy upto $10^{23}$ eV. Remembering that a part of this energy will be in the form of the rest energy of the emitted particles, the figure $10^{20}$ eV for the highest energy of the cosmic rays observed on Earth is readily comprehensible. Thus we see that the Holeums are a natural source of cosmic rays. The origin of the highest energy CRs is one of the outstanding puzzles of astrophysics [2-6]. Until now attempts have been made to see how the particles of cosmic rays can be accelerated to such high energies. To this end super super novae and pulsars [6], galaxies in collision [5] and giant relativistic jets emitted from the centres of active galaxies have been suggested as the sources of the cosmic rays. But these are speculations as there is no direct evidence that CRs are really emitted by them. Now we see that CR particles are created with such high energies that the need for exotic sources of acceleration does not arise.

IV BLACK HOLEUMS, INFLATION, BARYON ASYMMETRY AND MATTER BACKGROUND

Substituting the numerical value of $\Delta$ from equation (15) into equation (22) we find that the region c is given by $0.9059\ m_P \leq m \leq 1.6452\ m_P$. It can be shown that all the ground states n=1 in this region are the Black Holeums BH. For each m the next several excited states can be either H or BH. The remaining states are all H. This is shown in Table I. In this region c, the gravity dominates. As the temperature declines, their number density n also declines and their average separation $d = n^{-1/3}$ increases. Below the unification energy of $10^{16}$ GeV the gravity decouples from the other unified interactions and so do H, BH, HH and the LH from the primordial brew. Of these the BHs are themselves black holes while the others have stable ground states. The expansion of the universe causes the BHs to get more and more isolated. The amended Hawking theorem now applies to these isolated BHs and they begin to explode. The more massive ones are the first to go followed by the less massive ones. For a brief moment the universe exploded into innumerable mini-bangs of thermal radiation of particles given off by the exploding BHs. This scenario is very reminiscent of the bigbang itself and therefore we will call it the bigbang - two (BB2). Until now the universe was expanding at a uniform rate. But now due to the BB2 there is a huge increase in its rate of expansion. This is the familiar Inflation of the universe, which was postulated by Guth and others, to explain a number of problems such as the horizon problem, the flatness problem, the entropy problem etc. But its dynamical origin has remained unknown so far. In our theory the BHs not only get formed but also remain intact as long as they are nonisolated from one another. The condition for isolation is $d \gg \lambda$ where the latter is the Compton wavelength of the BH and d is the average separation already defined above. But the moment the gravity decouples from the other interactions and the expansion of the universe isolates them from one another the amended Hawking theorem begins to destroy them. These explosions give rise to the Inflation of the universe. The particles of the thermal radiation will collide with one another as well as with those of the primordial brew. And after a sufficiently long time the thermal equilibrium will be reestablished and the universe will resume uniform expansion albeit with an enhanced rate. The problems referred to above may now be addressed hopefully successfully.

In the standard cosmology the Inflation is placed after the GUT (Grand Unified Theory) epoch in order to produce the baryon asymmetry. In our theory also the Inflation comes out after the unification epoch because of the decoupling of the gravity at that epoch. The thermal nature of the BB2 provides the baryon asymmetry. This is because the Hawking radiation does not respect the CP symmetry. We note that all Holeums, including the BHs, were produced copiously and the region c is the second largest of a, b, c. Therefore this thermal radiation from BB2 may form a substantial component of the visible matter in the universe today. The BB2 would have produced GWs, too. They would form a part of the background of GWs in the universe today with a red shift of about $10^{29}$.

V HYPER HOLEUMS AND ULTRA HIGH ENERGY COSMIC RAYS

Numerically the region a, equation (23), is given by 1.6452 $m_P$ < m < 1.6818 $m_P$. Now the Planck mass, equation (2), contains ℏ and G and thus signals the onset of Quantum Gravity (QG). The bound state problem in the region a belongs to the strong field regime; that is, to General Relativity (GR). The moot point is whether the present treatment based on NG and the Schrodinger equation is adequate for this region. This point is discussed exhaustively in reference [1]. To within a factor of ten the answer is in the affirmative. It is based on the newly discovered Bound State Wisdom (BSW) which is culled from three layers of matter: atomic, nuclear and particle. It states that the asymptotic behaviour of the binding potential, as r→∞, determines the values of the bound state parameters to within a factor of ten; the actual behaviour of the potential within and near the bound state as well as the finite size of the constituents being of no consequence. In our case this means that we may replace GR by NG which is the r→∞ limit of the former and yet get an order of magnitude correct description of the region a. The latter, defined by equation (23), is divided into four equal parts $a_{1/4}$, $a_{1/2}$, $a_{3/4}$, $a_+$ and the results of the calculations are presented in Table I. As already mentioned above, the HHs have stable ground states but their n=2 and 3 states are BHs. The masses of the latter are shown in the parentheses in units of $m_P$ in Table I. When the rate of the gravitational interactions fell below the rate of expansion of the universe the HHs decoupled from the primordial brew. And because of the rarity of the collisions their number was frozen. Since they have stable bound states they will be present in the universe today. H and HH emit gravitational radiation when they make a transition from a higher state n' to a lower one n. But due to the lack of the GW detectors Hs and HHs are undetectable now. Because of the copious production of PBHs in the early universe the Holeums H and HH may form an important component of the DM. The HHs are the relics of the QG epoch. Although the collisions among the HHs are very rare, if such a collision takes an HH into one of its n=2 or 3 BH states, then the Hawking evaporation will destroy it instantly giving off a large number of particles. These are the ultra high-energy CRs. These events are reminiscent of the Centurian Events reportedly observed by the Brazil-Japan cosmic ray collaboration in the early seventies. If we could determine the total energy E released in such an event then according to the present analysis E = $M_n c^2$ where $M_n$ is given by equations (10) and (3). Knowing E one can determine m and n by referring to the Table I. This gives us direct information about the QG epoch. From Table I we see that such events, which we may call mini-bang events, would have their total energies E within the band 2.5 $m_P c^2$ and 3.0 $m_P c^2$, that is, within $3.0\times10^{28}$ eV and $3.6\times10^{28}$ eV. By comparison, at present the highest detected energy of CRs is about $10^{20}$ eV. These events are rare. For example, the cosmic ray particles reach the top of the Earth's atmosphere at rates ranging from one per square meter per century above $10^{17}$ eV to one per square kilometer per century above $10^{20}$ eV. The Auger project uses fluorescence detectors and couples them with a giant array of particle detectors covering 3000 square kilometers. It may be able to detect the ultra-QG band [7].

VI. LUX HOLEUMS AND DARK ENERGY.

A Lux Holeum has a stable ground state n=1. Therefore it must be present in the universe today. It is a bundle of gravitational energy. But it is not the graviton because an LH is a bound state of two black holes having an infinite tower of excited states. A graviton, on the other hand, is an elementary particle having no excited states. In the early universe the number of LHs would have been frozen at the instant when the rate of gravitational interactions fell below that of the expansion of the universe. Thereafter it did not change due to the rarity of collisions except in the following way. From the Table I we see that the n=2 and 3 states of an LH are BHs. If a rare collision takes an LH into one of its BH states it will burst into a large number of particles as already discussed above. These particles will be a part of the ultra-QG band of the CRs. The LH as well as the other Holeums cannot be detected today because of the lack of GW detectors. Therefore the LHs will form a part of the DE in the universe today [8-10]. The energy carried by an LH is greater than $2\times8^{1/4} m_P c^2$ = $4.10693\times10^{28}$ eV or 6.4 billion Joules! It could create a Tunguska –Siberia- type event. But the probability of interaction with matter is so low that we need not worry about getting hit. For example, a neutrino can go through the Earth without an interaction. And the gravitational interaction is far weaker than the weak one. The LHs were produced copiously during the QG epoch. Hence the LHs may make up an important component of the DE in the universe today.

A Lux Holeum and its tower of excited states is a unique quantum system containing matter-energy oscillations in it. This tower of quantum states is very similar to the pyrgons found in the higher dimensional Kaluza-Klein theories which have exactly the same composition with massless ground states and massive excited states. In the Kaluza-Klein theories the standard model particles live in the four-dimensional brane world whereas the pyrgons live in the bulk space made up of the compact extra dimensions. Our Lux Holeum and its excited states, though pyrgon-like, live in the in the four dimensional space-time.

VII. DARK MATTER AND DOMAIN WALLS.

By the time the particles of ordinary matter such as atoms and molecules were produced in the early universe the gravitational interaction had already become the weakest. In addition to the gravitational interaction the particles of ordinary matter had three other, stronger, interactions. This resulted in a segregation of H and HH from the particles of ordinary matter. But due to gravity the former still clung to the latter. This resulted in accumulation of the Hs and the HHs in the haloes of galaxies and in the domain walls. Since the Hs and the HHs are undetectable at present so are the haloes of the galaxies and the DWs. There is another reason for the accumulation of the Hs and the HHs in the haloes. This is the centrifugal force of rotating galaxies. The Holeums are generally much more massive than the particles of ordinary matter. Therefore the centrifugal force on them is greater. The greater concentration of the Hs and the HHs in the haloes of the galaxies will lead to greater emission of CRs from the haloes than from the discs due to pressure-ionization as discussed above. This is strongly supported by the conclusion: "The observed anisotropies can be accounted for only if the diffusion in the disc is much smaller than that in the halo" [4]. There is a vast literature on the DWs [11-13]. But it relates to the cause and the structure but not to their content. Here we are suggesting that the Hs and the HHs may form an important component of the DWs. Since the GHs consist of Hs and the HHs, the latter two would lead to the formation of Holeum stars in the GHs. And the Holeum stars would continuously emit not only the GWs but also the infra-QG band of CRs due to the pressure-ionization of Hs and the ultra-QG band of CRs due to the exploding n=2 and n=3 excited states of the HHs.

VIII. GRAVITATIONAL WAVES.

The energy spectrum of an individual Holeum H or an HH is formally identical with that of a hydrogen atom. The frequency $\nu_{n'n}$ of the gravitational radiation emitted by a Holeum when it makes a transition from a higher state n' to a lower one n is given by

$$\nu_{n'n} = \nu_0 (m/m_P)^5 (n^{-2} - n'^{-2}) \tag{27}$$

where $\nu_0 = m_P c^2/4h$ (28).

Here m varies from 0 to $m_c$ in the region b. Thus the Holeums H populating the latter will emit a continuous band of frequencies :

$$0 \text{ to } 4.525442 \times 10^{41} \text{ Hz} \tag{29}$$

We will call it the infra-QG band. The Hyper Holeums of the region a will give rise to another band called the ultra-QG band of frequencies given by

$$8.939355 \times 10^{42} \text{ Hz to } 9.978206 \times 10^{42} \text{ Hz} \tag{30}$$

The gravitational radiation emitted by the exploding BHs at the energy of $10^{16}$ GeV in the region c is of an entirely different type. It will form a cosmic GR back-ground with a red shift of about $10^{29}$. These are the parameter-free predictions of the theory. We recall that this is the "first glimpse" solution of the quantized gravitational bound state problem. The scale-invariant analysis of the Paradigm of the Truncated Hydrogen (POTHA) of reference I indicates that the energies and the frequencies in this model are given correctly to within a factor of the order unity and that the distances in this model are given correctly to within a factor of several. But the overall limit on accuracy is set by the BSW which indicates an accuracy of a factor of ten.

The cosmic microwave back-ground (CMB) experiment WMAP has revealed the following matter-energy content of the universe: 4% visible matter, 23% Cold Dark Matter, 73% DE.

The most popular candidates for DM are related to the SuperSymmetry (SUSY) [14]. These are the Lightest SuperSymmetric Particle (LSP), the Weakly Interacting Massive Particles (WIMP), the neutralino, the axion, etc. Experiments have been going on for the past several years. But none has been detected firmly as yet. At present DM searches are based entirely on SUSY particles. As far as the DM is concerned SUSY merely begs the question [15]. Besides the detection of SUSY particles would not in any way preclude Holeums and the others.

IX. DISCUSSIONS AND CONCLUSIONS

This is the Standard Paradigm of Quantum Gravity (SPQG). Its pedigree includes Bohr and Glashow. And a SP must be considered mandatory for every new phenomenon that we face. For example, non-standard-model phenomena are routinely assessed using the standard model calculations. Because it is rooted in past successes it never fails to produce a number of successes. And even where it fails it instructs. But before we list its many plausible successes we must discuss the challenges it faces. SPQG faces challenges from two commonly held beliefs. (1) All the PBHs got evaporated away in the early universe. This follows from the Hawking's theorem [16]. (2) QG implies a minimum length [17]. Particles of smaller size cannot exist. To take the first challenge, we have already made it plausible that the Hawking's theorem is valid only for isolated black holes and there were no isolated black holes in the early universe above the unification temperature. We have also shown above that the theorem is in conflict with the PND. The minimum length challenge is more serious. It may rule out the sub-Planck-mass Holeums H but it will not affect the BH, HH and LH. That is, Inflation, BA and ultra-QG CRs and ultra-QG GWs will not be affected. But infra-QG CRs and infra-QG GWs will be ruled out. However, the last word on QG has not been said. It is for this reason, for example, that the judgment on the ultimate fate of an evaporating black hole-whether it will end in a naked singularity, a Planck-mass stable relic or a complete burn-up- has been suspended. Hence it will be prudent to treat the sub-Planck-mass Holeums H as open to experimental verification. This is especially important since there is such a dearth of concrete theoretical predictions. And as already shown, the Hs are responsible for a wide range of phenomena such as DM, DW, GH, CRs and GWs and the prediction that the CRs originate in the GHs already has a strong experimental support. Most other DM candidates are almost self-serving constructs without providing any synthesis among the many enigmas of cosmology. In Kaluza-Klein theories with two extra dimensions, compactified at the millimeter scale, the Planck energy scale comes down to the TeV scale. In these models the Standard Model (SM) fields with we are familiar are constrained to live in a four-dimensional submanifold, called a brane, of the higher-dimensional space-time. And the non-standard model fields are constrained to propagate in the extra dimensions. The gravitons can access both the brane and the extra dimensions but at low energies they are confined mainly on the brane. Not only that, the black holes Hawking-radiate mainly on the brane. In this scenario, therefore, PBHs of mass as low as a TeV are viable. And so would their bound states, the Hs. The summary of this paper and that of reference [1] is as follows: (1) We predict stable Holeums H, HH and LH and the ephemeral BH. (2) Holeums H and HH obey a purely quantum mechanical exclusion principle and occupy space just like the particles of ordinary matter. (3) H and HH have a natural segregation property that puts them in the GHs and the DWs, if the latter exist. They constitute an important component of the dark matter in the universe. (4) The GHs are predicted to contain Holeum stars that emit the infra-QG and the ultra-QG bands of GWs and CRs. (5) The LHs constitute an important component of the DE in the universe. (6) We provide a plausible resolution of the enigmas of DM, DE, DWs, GHs, Inflation, BA, CRs and the role enigma of the PBHs in terms of Holeums of various types. (7) We predict the infra-QG and the ultra-QG bands of gravitational waves and their emission frequencies. (8) We also predict a cosmic background of GWs with a red shift of about $10^{29}$. (9) We also predict that far more CRs will come from the haloes of the galaxies than from their discs. This is already confirmed by the existing observations. (10) We predict presently invisible DE in the form of Lux Holeums carrying energy greater than 6.4 billion Joules. (11) We report a unique quantum system containing matter-energy oscillations similar to the pyrgons in the Kaluza-Klein theories except that our system lives in the observable four-dimensional brane-world.

# REFERENCES

(1) Chavda L.K. and Chavda Abhijit L. 2002.Class. and Quantum Gravity **19**,2927; hereafter referred to as the reference I.
(2) Teller E. 1954. Rep. Prog. Phys.**17**, 154.
(3) Burbridge G.R. and Hoyle F. 1964. Proc. Phys. Soc. **84**,141.
(4) Giler M. 1983. J. Phys. G: Nucl. Phys. **9**, 1139.
(5) Al-Dargazelli S.S., Wolfendale A. W., Smialkovsky A. and Wdowczyk J. 1996.J.Phys. G: Nucl. Part. Phys. **22**,1825.
(6) Erlykin A.D., Wibig T. and Wolfendale A.W.2001.New J. Phys. **3**, 18.
(7) http://www.physics.adelaide.edu.au/astrophysics/index.html.
(8) Morison, I. 2003.Phys.Edu.38,205; and references contained therein.
(9) Frampton P.H. 2002.A.I.P. Conf. Proc.**624**, 59.
(10) Neves R. and Vaz C. 2002.Phys. Rev. **D66**, 124002.
(11) Artstein Y., Sonnenschein J., and Kaplunovsky V. S. 2001. J. High Energy Phys. **02**, 040; and references contained therein.
(12) Kallosh R., Prokushkin S. and Shmakova M. 2001. J. High Energy Phys. **07**, 23.
(13) Forbes M. M. and Zhitnitsky A. R.2001. J. High Energy Phys. **10**,013.
(14) Arnowitt R., Dutta B. and Santoso Y.2002.Phys. At. Nuclei. **65**,2218.
(15) Pretzi K. July 2000. New J. Phys. **2**.
(16) Hawking S.W. 1974.Nature **248**,30.
Hawking S.W. 1975.Commun. Math. Phys. **43**, 199.
(17) Garay L.J. 1995.Int.J.Mod.Phys.**A10**,145.

Table I

| n | b | c | $a_{1/4}$ | $a_{1/2}$ | $A_{3/4}$ | $a_+$ |
|---|---|---|---|---|---|---|
| 1 | H | BH | HH | HH | HH | LH |
| 2 | H | H/BH | BH(2.5342) | BH(2.5308) | BH(2.5270) | BH(2.5227) |
| 3 | H | H/BH | BH(2.9646) | BH(2.9732) | BH(2.9817) | BH(2.9899) |
| 4 | H | H | HH | HH | HH | HH |
| 5 | H | H | HH | HH | HH | HH |

TABLE CAPTION

Table I. The class-structure of Holeums. See equations (20)-(24) for the definitions of the regions a, b, c. The masses of the BHs in the region a with n=2, 3 are given in parentheses in units of $m_P$.